**RESEARCH ARTICLE**

JRST | WILEY

# Navigating socio-emotional risk through comfort-building in physics teacher professional development: A case study

**Maggie S. Mahmood**  | **Hamideh Talafian**  |
**Devyn Shafer**  | **Eric Kuo**  | **Morten Lundsgaard**  |
**Tim Stelzer** 

Department of Physics, Department of Curriculum and Instruction, University of Illinois Urbana-Champaign, Champaign, Illionois, USA

**Correspondence**
Maggie S. Mahmood, Department of Physics, University of Illinois Urbana-Champaign, 1101 W. Green Street, Champaign, IL, USA.
Email: maggiesm@illinois.edu

**Funding information**
National Science Foundation, Grant/Award Number: DRL-2010188

**Abstract**

In teacher professional development (PD), grouping teachers with varying levels of experience can be a productive and empowering way to stimulate the exchange and co-generation of content and pedagogical knowledge. However, less experienced teachers can face socio-emotional risks when engaging in collaborative science content reasoning tasks with more experienced colleagues, and these risks may impact the collaborative experience of both parties and the learning environment in teacher PD. This exploratory case study examines the process of productively navigating socio-emotional risks and interpersonal tensions encountered by a veteran and pre-service physics teacher during one episode of discussing physics content. We use a single term, *comfort-building*, to encapsulate discursive moves that result in increased feelings of comfort and safety by the participants. Comfort-building includes moves that serve to mitigate social risk, ease tension, and avoid discomfort, as well as those geared toward finding common ground and co-navigating challenges. These moves can carve out conversational space for teachers to more confidently face risks associated with being accountable to the physics content knowledge and engage in discipline-based conversations more deeply. The presented case was followed by video-stimulated individual interviews to determine how consciously the teachers connected their participation to explicit risk and comfort. This case study highlights an affective dimension for consideration in the continued study and facilitation of science teacher PD, especially programs that bring together teachers with a variety of backgrounds and skill sets.

## 1 | INTRODUCTION

Near the end of her first year of teaching high school physics, Jessica reflected on the initial risks and eventual comfort she experienced when discussing physics content with a more veteran teacher in a professional development (PD) activity:

> I remember when I first started reading [the prompt] that I did not remember anything about kinetic energy to be honest, and I got a little worried that I was gonna look stupid in front of everybody [...] I remember at the moment we got the question I was like, *do we have to answer this? 'cause I don't know.* But then we just talked about how we felt about [the question], and I was like, *okay, this is easy, I can do that* [....]

Jessica also explained how her activity partner, Lisa, a physics teacher with 25 years of experience, helped her gain confidence. Jessica said of the experience with Lisa:

> [Lisa] also definitely never told me I was wrong about anything. She actually – if I had an idea and it was, like, a half-assed idea, she would definitely go with it and fully bring it out of me, and so that made me feel more confident. [...] I had finally wrapped my head a little bit more around high school physics by the end [...] and she wasn't just leading at that point. You can tell I was like, *no, I have my own ideas now.*

Here, Jessica reports achieving comfort and increased ability to contribute to a collaborative physics content reasoning task in the face of the socio-emotional risks that can arise when discussing science content with a more experienced peer. However, this may not be the case for other science teachers during similar collaborations in content-focused teacher PD activities. The *socio-emotional risks* of looking unknowledgeable or incompetent in front of peers during teacher PD activities can lead teachers to avoid revealing their thinking and (mis)understanding, stymieing potential learning benefits of collaborative PD activities (Finkelstein et al., 2019; Koellner et al., 2007; Vedder-Weiss et al., 2019). In this way, efforts to deepen teachers' content knowledge in PD programs may fail if the risks of participation are too high. Furthermore, PD programs that bring together teachers with diverse teaching and content knowledge expertise may further magnify these risks of embarrassment when novice teachers are paired with more knowledgeable veteran teachers. Considering this challenge of teacher PD, this article focuses on *comfort* and *socio-emotional risks* in secondary science teacher PD activities.

Recent work on learning has shifted away from studying single traits of learners (e.g., motivation, interest, etc.), toward a more holistic theory of contextualized, affect-driven conceptual learning that occurs at the intersection of affective, cognitive, and metacognitive dimensions of learning (Efklides, 2017; Jaber et al., 2023). In this new approach to studying learning, affect is not considered a separate construct representing a state or trait; instead, it emerges as a fluid, dynamic construct shaped within the learning context (Vea, 2020). Hence, in response to recent works that have called for greater attention to the contextualized role of *affective phenomena* in learning (Avraamidou, 2020; Curnow & Vea, 2020; El Halwany et al., 2021; Lanouette, 2022), this study illustrates how affect and cognition are intertwined and how supporting one aspect can facilitate progress in the other.

This exploratory case study examines how, in the face of socio-emotional risks posed by a Content Knowledge for Teaching Energy (CKT-E) task (Etkina et al., 2018), a pre-service teacher (Jessica) and a veteran teacher (Lisa) created a safe space to practice and develop their CKT-E through comfort-building moves. We use video data of their PD interaction and post-episode interviews with the two teachers to investigate two questions:

1. How can socio-emotional risks related to science content surface in secondary science teacher PD discussions?
2. What conversational, comfort-building moves can teachers use to mitigate these risks and support a safe space for discussing scientific content and reasoning?

This study contributes to the body of literature in two significant ways. First, by focusing on the affective dimension of learning within the interactions of two teachers, this case study informs our understanding of knowledge co-construction when socio-emotional risks are involved. Second, this study offers theoretical insights into how teachers mitigate risk through a set of comfort-building moves, which could lead to potential practical applications in teacher PD catering to teachers with varying experience levels.

## 2 | BACKGROUND AND THEORETICAL FRAMEWORK

### 2.1 | Socio-emotional challenges of teacher–teacher interaction in PD settings

Teacher PD programs commonly leverage teacher–teacher interaction as a mechanism for developing content and pedagogical knowledge. Sociocultural mechanisms of teacher interaction are a driving force for transformative outcomes, such as the development of teacher identity and self-efficacy (Barab et al., 2002; Eshchar-Netz & Vedder-Weiss, 2021; Finkelstein et al., 2019; Little, 2002; Takahashi, 2011). Yet, teachers can also face interactional challenges due to power and knowledge imbalances in the group that impede their engagement. Hierarchical relationships between novice and veteran teachers can develop (Barton & Tusting, 2005; Eshchar-Netz & Vedder-Weiss, 2021; Horn & Little, 2010), leading to veterans dominating conversations (Liu, 2013), taking the role of "helper" (Goh & Fang, 2017) and guiding novice teachers (Horn & Little, 2010). Via this positional imbalance, novice teachers may become hesitant to share their uncertainties (Costache et al., 2019) and disengage from the group (Barton & Tusting, 2005).

Along with power and positioning, this interactional challenge for novice and veteran teachers can be affective and emotional as well. Affect and emotion are integrally related with social and relational dynamics in learning contexts (Jaber et al., 2023; Vea, 2020). The interplay between emotions, power, and positionality been described theoretically as a complex web of discursive practices (Abu-Lughod & Lutz, 1990) where power dynamics and culture play a role in shaping emotions (Cornell, 2000; Holstein & Gubrium, 2000). In this view, emotions are social constructions which shape one's values, ideas, and practices (Rivera Maulucci, 2013; Rose, 1990), are shaped by power relations (Foucault, 1977), and are deeply intertwined with cognition (Keltchermans, 2005; Rivera Maulucci, 2013).

For teachers, positive and negative emotional experiences in PD contexts can impact their learning and professional identities. Broadly, teachers' emotions, self-perceptions, and positionality can shape their professional identities (e.g., Moore, 2008; Nias, 1989), relationships with stakeholders in a school setting (Blackmore, 1996), and ultimately their pedagogical practices (Zembylas, 2004). Prior studies have shown that socio-emotional threats in formal PD contexts can undermine teacher engagement and learning. Teachers can experience threat associated with exposing weaknesses in their knowledge or teaching practice to their peers (Finkelstein et al., 2019; Koellner et al., 2007; Vedder-Weiss et al., 2019). Novice teachers, in particular, may become reticent to share ideas and risk "looking stupid" in front of more experienced and knowledgeable veteran teachers (Charteris & Smardon, 2015). Considering such obstacles to teacher learning in PD contexts, this paper explores the interactional work teachers can do to mitigate socio-emotional risks in PD settings.

## 2.2 | Mitigating socio-emotional risks through comfort-building moves

In this study, we cast comfort as a key ingredient in teachers' ability to develop disciplinary proficiency and articulate the construct of *comfort-building* to describe the interactional work that groups do to manage socio-emotional risks. Research on collaborative dynamics has documented the ways in which groups navigate tensions during problem-solving activities (Andriessen et al., 2013; Appleby et al., 2021; Asterhan, 2013; Barron, 2000, 2003; Berland & Reiser, 2011; Conlin & Scherr, 2018; De Dreu & Weingart, 2003; Goffman, 1955; Jaber, 2015; Johnson & Johnson, 1979; Sohr et al., 2018; Zhang et al., 2021). The moves described in these studies can be classed into two types: (a) moves that place distance between participants and their ideas, and (b) expressions of empathy in the sensemaking process and of care for others' professional growth. In this paper, comfort-building moves describe all discursive moves that result in feelings of comfort and safety by the participants.

The first set of comfort-building moves create distance between the participant and their ideas to mitigate risk and tension. *Facework* describes discursive attempts by individuals to save or restore face when one perceives a threat to their social image or status during a group interaction (i.e., a face-threat) (Goffman, 1955). Facework can have a "defensive orientation toward saving [one's own] face" but also a "protective orientation toward saving the other's face" (Goffman, 1955, p. 217). Facework has been documented in teacher PD interactions (Goffman, 1955; Vedder-Weiss et al., 2019). Like facework, *epistemic distancing* (Conlin & Scherr, 2018) describes how speakers can soften their epistemic stance through actions like joking, quoting, or hedging to create distance between themselves and their claims to mitigate potential face threats when presenting or critiquing ideas in a group. Unlike facework, which is largely reactionary in the wake of tension (face threat), epistemic distancing is prospective, reducing one's commitment to an idea during its initial presentation. Similarly, other studies have demonstrated how humor (Conlin, 2012) and playful talk (Sullivan & Wilson, 2015) can help groups navigate tensions. In some cases, tensions build so high that group members decide to disengage from or end the interaction. The term *escape hatch* refers to "a set of discourse moves through which participants close the conversational topic, thereby relieving tension, but before a conceptual resolution is achieved" (Sohr et al., 2018, p. 883). Like facework, taking an escape hatch in group problem solving is often reactionary to tensions arising in a group due to uncomfortable and seemingly unresolvable disagreements.

Whereas the above comfort-building moves are responses to realized or potential socio-emotional tensions and risks, a second class of moves motivated by participant empathy and care can support the building of social rapport and trust, ultimately allowing a group to work further into potentially risky activities before calling them off. Appleby et al. (2021) argue that *social caring* expressed outside of the classroom can promote students' disciplinary engagement in the classroom. More generally, activities that might be considered "off-task" can contribute to productive collaboration (e.g., Langer-Osuna et al., 2020), and moves that communicate and reaffirm social caring can be one type of productive "off-task" activity. Empathy and care can also be tied to a group's collaborative knowledge construction activity. *Epistemic empathy* (Jaber, 2015; 2021; Jaber et al., 2022) is the humanizing awareness of the value of another person's ideas, even when they conflict with canonical knowledge. Jaber et al. (2018) characterize the ways that pre-service teachers can express epistemic empathy for their students' scientific reasoning, which included appreciating epistemic affect, finding merits and value in their ideas, and expressing curiosity and interest in those ideas. Similarly, Krist and Suárez (2018) argue that science students can express epistemic care for others' thinking by asking them questions, entertaining ideas, following their logical implications before evaluating as right or wrong, and using data to invite conversation.

## 2.3 | The role of safe space in facilitating comfort-building moves

Establishing a safe space is a key process in building comfort in group problem solving tasks, which can in turn facilitate the freer exchange of ideas and the co-development of disciplinary knowledge. The term *safe space* denotes spaces in which individuals can support one another's learning (Blair, 2008; Holley & Steiner, 2005; Lave & Wenger, 1991; Rogoff, 1990; Wenger, 1998). For teachers' professional learning, a safe space is an environment where they can interact with one another and navigate socio-emotional tensions while being challenged to approach their discipline in new ways (Remillard & Geist, 2002) without being impeded by hard feelings or unnecessary pressure (Gayle et al., 2013; Holley & Steiner, 2005). Mutual respect and trust are integral components of a safe space where individuals can take risks and share their honest opinions without fear of making mistakes or being humiliated (Gayle et al., 2013; Turner & Braine, 2015). Relational trust can help teachers be better learners and develop relationships with other teachers (Blair, 2008; Borko, 2004; Sparks, 2002). When teachers experience trust, encouragement, and support from their community, they are more comfortable trying new teaching practices (Bryk & Schneider, 2002; Wanless et al., 2013).

The co-creation of a safe space by the participants in this study can be connected to these two classes of comfort-building strategies seen in the literature: (a) distancing self from disciplinary ideas, and (b) expressing empathy and care in collaborative activities. Although much of the prior research related to comfort building has focused on students (and their science-related sense-making discussions, in particular), similar comfort-building moves may be used by teachers to construct a safe space for learning in PD settings. This article provides a case study of one such comfort-building interaction between teachers, illustrating some of the ways novice and veteran teachers can navigate socio-emotional risk while participating in a teacher PD program.

# 3 | METHODS

## 3.1 | Research context: A high school physics teacher professional development program

This case study focuses on an episode within a Midwestern US physics teacher PD partnership aiming to develop a high school physics teacher community of practice (CoP) (Lave & Wenger, 1991; Wenger, 1998). The authors, who are members of a large, public research university, organize and facilitate the program, which encourages teachers to adopt new curricular materials and pedagogies from other teachers and a university-level curriculum. The program's goal is for teachers to share, not just take up, teaching materials with each other. The university team positions itself as a peer and supports teachers in selecting, adapting, and implementing tools that fit their unique teaching contexts. During two 4-day-long summer workshops and online weekly meetings, teachers develop instructional materials, as well as their physics content and pedagogical knowledge. At the time of the episode, the group was comprised of four experienced physics teachers, ranging from 10 to 25 years of teaching experience, and one pre-service physics teacher, who had joined the group to get support before the start of her first semester of high school teaching.

## 3.2 | Task context: The CKT-E assessment

This case study focuses on two teachers discussing the answers to physics questions from the CKT-E assessment (Etkina et al., 2018), an activity occurring on the fourth day of the second summer workshop.

### 3.2.1 | CKT-E in physics teacher professional development

*Content Knowledge for Teaching* (CKT) encapsulates an integrated set of teacher disciplinary, pedagogical, contextual, and cultural know-how that allows teachers to catalyze their disciplinary content knowledge into meaningful learning experiences for students (Ball et al., 2008; for related discussion of Pedagogical Content Knowledge, see Shulman, 1986). Said another way, CKT is the knowledge teachers use in disciplinary "tasks of teaching" (Ball, 2000), such as evaluating and responding to student thinking in-the-moment or adapting existing instructional materials to fit one's classroom context. In addition to the "common" content knowledge required to demonstrate competence in typical performance tasks, CKT encompasses more specialized content knowledge and pedagogical content knowledge that facilitates the unpacking of disciplinary ideas often called for in teaching but not on typical performance tasks (Ball et al., 2008; Krauss et al., 2008).

In physics teacher PD, there has been a call for teacher training to attend to physics-specific pedagogy, including planning for and responding to typical patterns in physics student reasoning (Meltzer et al., 2012; Meltzer & Otero, 2015; Sunal et al., 2019). This discipline-specific PD effort has included the development of assessments to evaluate the state of physics teachers' CKT for different physics topics (see Gitomer & Bell, 2013; Kirschner et al., 2016). Robertson et al. (2017) developed a method of identifying physics-specific CKT from in-classroom interactions of physics teachers teaching energy, and this inspired the creation of a multiple-choice CKT-E assessment (Etkina et al., 2018). These efforts are supported by research suggesting a connection between teachers' development of CKT and the quality of their instructional explanations of physics phenomena (e.g., Kulgemeyer & Riese, 2018).

### 3.2.2 | Focal questions: Two blocks Q1 and Q2

The CKT-E assessment items discussed in this study are Two Blocks Q1 (Table 1) and Q2 (Table 2). Each question has four multiple-choice options: one correct answer and three representing possible student errors requiring refinement of physics knowledge and how it is applied. Because the scenarios cover topics taught in high school physics, one might predict teachers would not find them challenging. However, in a survey of 220 US high school physics teachers, only 31% answered Q1 correctly and 62% answered Q2 correctly (Etkina et al., 2018).

Two Blocks Q1 and Q2 are designed to test teacher understanding of two key physical principles by embedding them in examples of student reasoning derived from a study of real physics classroom scenarios. First, teachers must understand that the work done on an object by the constant force of the air blower is equal to the magnitude of the force times the distance the object traveled (mathematically, this can be written as $W = Fd$, where $W$ is the work done on the block, $F$ is the magnitude of the force of the air blower, and $d$ is the distance traveled). Because both blocks experience the same constant force and travel the same distance, the work done on each block is the same. Second, the work-energy theorem states that the change in kinetic energy (KE) of an object is equal to the work done on it (which can be written as $W = \Delta KE$, where $\Delta KE$ is the change in KE). Because both blocks have the same amount of work done on them, they also experience the same change in KE. Because the blocks both start with the same KE at rest (zero), experiencing the same change in KE means they must have the same final KE. Tables 1 and 2 summarize the CKT-E required for solving Two Blocks Q1 and Q2, and a more in-depth treatment of the significance of each answer choice for Q1 and Q2 is provided as they arise in the following presentation of the case.

**TABLE 1**  CKT-E Question 1 (Two Blocks Q1) discussed in the teacher interaction (from Etkina et al., 2018) with key points for CKT-E reasoning.

Mr. Andreou's class is in the middle of discussing possible factors that determine the change in kinetic energy of objects. Students have collected data in the following experimental setup.

Two blocks with different masses are free to slide on a very, very smooth table between two parallel lines. An air blower pushes each block horizontally, exerting the same constant force. Both blocks start from rest and cover the same distance on their track under the action of the air blower. The experimental data collected by the groups support the claim that the final kinetic energies of the two blocks are equal.

1. Of the following four student responses, which is the most correct and complete account of the kinetic energies of the two blocks being equal?

| Choice | Student response | Key points | Pilot response (%) |
|---|---|---|---|
| A | The two blocks have equal final kinetic energies because the blower transfers to each of them equal amounts of energy per second | **Incomplete:** Incorrect application of the impulse-momentum theorem, related to how forces affect systems as they act over *time* rather than over *distance* | 11 |
| B | The two blocks have equal final kinetic energies because the higher final speed of the lighter block compensates for its smaller mass | **Incomplete:** Uses compensatory reasoning to describe how two blocks with different masses *can* have the same KE; does not provide a principled explanation for why the KEs *must* be the same. Less massive block must have a greater speed if the quantity $\frac{1}{2}mv^2$ is the same for both blocks. Description of changes in mass and speed offsetting does not *explain* why quantitative effects of higher speed and lower mass for the lighter block must exactly offset to produce equal KE to the heavier block | 46 |
| **C** | The two blocks have equal final kinetic energies because the blower transfers to each of them equal amounts of energy per meter | **Most correct:** If both blocks experience the same applied force, each meter traveled results in the same amount of work done, so the same amount of kinetic energy (KE) is imparted to each block per meter traveled | 31 |
| D | The two blocks have equal final kinetic energies because when there is negligible friction, mechanical energy stays constant | **Incomplete:** Contains an error related to another version of work-energy theorem: the amount of work done by *non-conservative* forces equals the change in total *mechanical* energy (TME), or the sum of KE and potential energy (PE) | 11 |

**TABLE 1** (Continued)

| Choice | Student response | Key points | Pilot response (%) |
|---|---|---|---|
| | | Fails to address that the blower does non-conservative work, producing the same increase in TME for both blocks. The blocks' PE does not change in this scenario, thus the increase in TME can be attributed to an increase in KE for both blocks | |

*Note*: Differing from the intent of the original assessment, teachers were prompted to not only attempt to answer the questions, but to discuss the pedagogical value of these questions within the context of a teacher professional development setting. The *Pilot Response* column represents the percentage of teachers selecting each response in the survey of US high school physics teachers conducted by Etkina et al. (2018).

### 3.2.3 | Collaborative implementation of CKT-E assessment items

Though the CKT-E was designed for *assessing* teacher knowledge, this case study takes place in a PD program context where the CKT-E items had been recast as a learning tool for *developing* physics teachers' CKT-E. The teaching scenario of Two Blocks Q1 and Q2 contextualizes teachers' collaborative discussion of content in the practice of teaching and toward interpreting and evaluating students' thinking, two features of high-quality PD (Borko et al., 2010). During this PD activity, the teachers were given a subset of problems from the CKT-E assessment and were asked to work through them collaboratively. PD facilitators hoped that teachers would discuss the incorrect answers in the questions, through which they would practice the teaching task of evaluating the merits and weaknesses of students' reasoning and potentially develop their own CKT. Teachers were also asked for feedback on the value of the questions for fostering productive teacher discussions and if they should be used in future PD activities. To our knowledge, this is the first documented case of the deployment of CKT-E assessment questions for collaborative discussion in teacher PD.

Throughout the PD design and implementation, we, as PD facilitators, explicitly attended to power dynamics and teacher affect, and these considerations directly implemented the facilitation of this activity. A central focus of our facilitation was challenging the view that we, a group of university physicists, are authorities on physics teaching. Instead, we explicitly positioned our role as facilitating the exchange of pedagogical ideas to help teachers identify new pedagogies that they were interested in exploring and developing. In the CKT-E discussion, the facilitator framed the task loosely and abstained from joining the conversation to allow teachers the chance to share their pedagogical thinking with each other. Another design goal of the CKT-E activity was to minimize the potential for the teachers—who have a wide range of backgrounds and physics teaching experience—to frame the activity as being assessed by university physicists. The collaborative activity, task framing of providing feedback on the value of these questions for future PD, and the teaching scenario context of the CKT-E questions were all employed to distance the activity from typical assessment contexts. This activity design was also informed by our attention to teacher (dis)comfort, as we believed teachers would be more comfortable with a discussion

**TABLE 2**   CKT-E Question 2 (two blocks Q2) discussed in the teacher interaction (from Etkina et al., 2018) with key points for CKT-E reasoning.

Mr. Andreou's class is in the middle of discussing possible factors that determine the change in kinetic energy of objects. Students have collected data in the following experimental setup.

Two blocks with different masses are free to slide on a very, very smooth table between two parallel lines. An air blower pushes each block horizontally, exerting the same constant force. Both blocks start from rest and cover the same distance on their track under the action of the air blower. The experimental data collected by the groups support the claim that the final kinetic energies of the two blocks are equal.

2.  A student's written explanation states, "The two blocks have equal final kinetic energies when they cross the finish line because the blower pushed each block equally hard." If the students were to use similar reasoning to compare the final kinetic energies for the two blocks in each of the variations of the experiment below, for which variation will the student's comparison of the final kinetic energies of the blocks be correct? Assume in each variation that the blocks start from rest.

| Choice | Student response | Key points | Pilot response (%) |
| --- | --- | --- | --- |
| A | The experiment is repeated on the same very, very smooth table. The blower pushes the blocks with the same constant force for the same time interval | **Incomplete:** Applies impulse-momentum theorem (forces over time) instead of work-energy theorem (forces over distance) | 17 |
| B | The experiment is repeated on a table that has small but not negligible friction. The blower exerts the same constant force on each block over the same distance | **Incomplete:** Friction is a mass-dependent force; the heavier block will experience a greater force of friction and, therefore, work done by friction, so the final kinetic energies will not be equal | 13 |
| C | The very, very smooth table is slanted upward; the blower exerts the same constant force on each block uphill parallel to the track over the same distance | **Incomplete:** Different changes in gravitational PE due to differences in block mass will, all else being equal, indicate that the final kinetic energies will not be equal | 8 |
| D | Instead of a blower, each block is pushed by the same compressed spring as the spring is released. The experiment takes place on the original very, very smooth table | **Most correct:** The same compressed spring will do the same amount of work (exerting the same average force over the same distance) on each block as it decompresses Also, the PE stored in the compressed spring is transferred into the KE of the block when the spring decompresses | 62 |

*Note*: Differing from the intent of the original assessment, teachers were prompted to not only attempt to answer the questions, but to discuss the pedagogical value of these questions within the context of a teacher professional development setting. The *Pilot Response* column represents the percentage of teachers selecting each response in the survey of US high school physics teachers conducted by Etkina et al. (2018).

activity than a written exercise. We were also intentional in selecting teacher pairs; we selected a partner for the sole preservice teacher who she seemed to get along with well socially.

## 3.3　|　Data collection and case selection

The primary data used in this case study are audio-video recordings of two teachers—veteran teacher Lisa and pre-service teacher Jessica (pseudonyms)—discussing Two Blocks Q1 and Q2. These data were collected during the partnership's summer PD meetings, which were conducted online through a videoconferencing program. The research team made recordings of the online meeting's main and breakout rooms during each session and acted as participant-observers in large group conversations, taking field notes throughout. The teachers consented to being recorded and were aware that the recording was taking place.

These observations led to the selection of a 1-h-long episode, out of approximately 80 h of video data from the PD workshops, as the basis for an exploratory case study (Yin, 2009) of socio-emotional risk and comfort building in a teacher PD setting. Our goal with this exploratory case study is to advance the theory of socio-emotional risk mitigation in-context—without suggesting that the particular interaction is representative, either for other individuals or for Lisa and Jessica in other times or contexts. We chose to focus on this episode because it well illustrates how one pair managed to ameliorate feelings of discomfort in approaching a problem and laid the groundwork for the pre-service teacher to shift from being the passenger to a driver in the interaction. Doing so opened up opportunities for teachers to discuss the problem content, practicing, and developing their CKT. Also, the veteran-novice pair highlights socio-emotional risks that newer teachers can face when discussing physics content and pedagogy with more experienced peers, a common activity of in-service PD workshops.

In addition, Lisa and Jessica were interviewed 7 and 9 months after the episode, respectively, using video-stimulated recall to solicit reflections on their experiences during the episode. The interviews lasted 1 h each. We use these teacher reflections to support our analytic interpretations of risk and comfort in the episode.

## 3.4　|　Participants

### 3.4.1　|　Lisa, a veteran teacher

Lisa is a veteran physics teacher with 25 years of experience. In a survey administered prior to her engagement with the program, Lisa reported that her overarching goal is getting students—who may be resistant at first—to love physics by honoring their agency and providing equitable instruction. Self-aware that her teaching style deviates from a "traditional path," she describes herself as a teacher who strives to fight the stereotype that physics is too hard to learn. Based on evidence we have of Lisa's teaching philosophy from surveys and other PD interactions, she sees her role as a teacher to build students' confidence and intrinsic interest in physics topics, and is therefore hesitant to use language like "incorrect" or "wrong" when referring to student ideas. She views herself as a lifelong learner who approaches each new experience with a goal of self-growth. This mindset is encapsulated in her classroom practice—evident in the ways in which she talks about student growth, the activities she creates for her classes, her grading philosophy to allow multiple revisions, and her encouragement to her students to adopt this same growth mentality. We view Lisa's emphasis on openness to learning and growing as foundational to the comfort building that occurs in the episode.

### 3.4.2 | Jessica, a pre-service teacher

At the time of the focal interaction, Jessica had recently completed a master's degree in physics and was yet to begin her first year of teaching at a small, private high school. During the PD workshop, Jessica was a more peripheral member of the CoP, attending meetings as needed to support her course planning. However, Jessica consistently asked questions of the veteran teachers and was generally confident, outspoken, and actively engaged in the PD activities that summer. Jessica did not have a formal background in education, but she had experience as a physics teaching assistant during her undergraduate and graduate studies, which ignited her decision to become a teacher. However, her most recent experiences as a teaching assistant for undergraduate courses had given her a distaste for pedagogy that focuses on drills, formulaic approaches, and an emphasis on getting the right answer above all else. In the PD program, she was open to learning new things and was excited about the opportunity to learn from more veteran teachers and get feedback on her ongoing work.

## 3.5 | Data analysis procedure

First, the selected hour-long episode was transcribed. In line with an interaction analysis methodology (Jordan & Henderson, 1995), the research team watched the video together twice, pausing as needed to discuss interpretations of notable events. After multiple team discussions, a focus on comfort building and managing socio-emotional risk crystallized. The first author then assembled an initial narrative-style analysis of the episode and narrative profiles of each teacher based on past interactions with the teachers as PD facilitator as well as teacher surveys. This analysis chunked the episode into distinct sections marked by shifts in the conversation, then looked for moves each teacher made that facilitated these shifts. The wider research team then reviewed the analysis and iterated several times until an initial set of key themes and discursive patterns emerged. Following this initial analysis, an author who had not been present at the PD workshop for this episode conducted individual video-stimulated recall interviews with the teachers to clarify teachers' experiences around this episode. Following feedback on an earlier version of this article, the case study narrative was revised to highlight how the comfort-building moves created space for teachers to practice and develop their CKT-E. The episode was originally transcribed using Jeffersonian transcription conventions (Jefferson, 2004) to document prosody, body language and tone. Additional methods are available in more detail as Supporting Information accompanying the article. For ease of reading, the transcript presented is simplified to the cues leveraged in the analysis.

To complement the in-depth analysis of teacher interaction, we created a graphic timeline of the thematic content of the teachers' talk. Three coders (the first, second and third authors) segmented the transcript by speaker turn number and created a thematic coding scheme through initial open coding, discussion, and development of a finalized, consensus coding scheme. Table 3 describes the final codes. Two of the original coders then coded the full episode in MAXQDA (VERBI Software, 2021), resolving any intercoder disagreements mid-way through and at the end of the coding process. Intercoder agreement before discussion was 92% (Cohen's $\kappa = 0.69$). This timeline visualizes a coarse, macroscopic view of the conversational patterns and themes in each episode, which we subsequently elaborate through the narrative analysis.

**TABLE 3**  Coding scheme categories for teacher interaction episode.

| Code | Title | Description |
|---|---|---|
| 1 | Disagreement | Used when teacher expresses a differing opinion, directly challenging the other teacher |
| 2 | Checking in | Characterized by an appeal to the other teacher to see if they are thinking similarly about something. Signaled by phrases, often questions, such as "right?" or "what do you think?" |
| 3 | Affirming the challenge | Statements that kindly acknowledge the task is difficult, specifically for one's partner |
| 4 | Physics content | Statements surrounding canonical physics content knowledge, whether in service of solving the presented problem, or not |
| 5 | Pedagogy | Statements surrounding physics teaching methods or in-classroom use of instructional materials |
| 6 | Agreeing | Verbal or physical evidence of agreement in any aspect of the conversation |
| 7 | Deemphasizing correct answer | Any speech that attempts to lessen the importance of the eventual selection of a correct answer within the activity; may include discussions of strengths of incorrect answers or reasons why it is important to hear out incorrect answers as a teacher |
| 8 | Personal experience | Insertion of an individual's life experience into the conversation; examples include discussion of one's experience as a student and as a teacher |
| 9 | Critiquing task | Pointing out weaknesses in the prompt itself, and/or motioning to steer toward a different epistemic activity than the one presented |
| 10 | Discomfort and vulnerability | Speech that indicates a teacher is (or at one point was) uncomfortable, confused, frustrated, apologetic, or lacking confidence |

# 4 | RESULTS

Our analysis suggests four types of comfort-building moves that can mitigate socio-emotional risks and contribute to a safe space allowing teachers to discuss physics content. These comfort-building moves are: (a) Challenging the Epistemic Activity of Finding the One Correct Answer (b) Revealing Vulnerability, (c) Collaboration-Centered Problem-Solving, and (d) Storytelling. These moves open space for the discussion of physics content on Two Blocks Q1 and Q2. Through this discussion, not only do Lisa and Jessica come to consensus around the correct answer choices and explanations, but, by discussing the incorrect multiple-choice options, they also practice and develop their CKT-E.

## 4.1 | Episode Section 1—Challenging the epistemic activity and revealing vulnerability

Section 1 of the episode (Turns 1–26) begins as Lisa and Jessica start discussing Two Blocks Q1 (Table 1). As both teachers are unfamiliar with the CKT-E format, turns 1–26 are characterized by Lisa and Jessica scanning through the answer choices and noticing that they each seem to have elements of a correct idea within them. Choice C gives the correct physical explanation by reformulating the work-energy theorem: if both blocks experience the same applied force, each meter traveled results in the same amount of work done, so the same amount of KE is imparted to each block per meter traveled. The remaining, incorrect choices represent plausible student errors. Distractor choice A represents an *incorrect application of the impulse-momentum theorem*, which relates to how forces affect systems acting over time rather than over distance. Distractor choice B employs *compensatory reasoning* to explain how the two blocks *can* have the same KE, but not why the KEs *must* be the same. Although the block with less mass must have greater speed if the quantity $\frac{1}{2}mv^2$ is the same for both blocks, choice B is not a principled explanation of why quantitative effects of higher speed and lower mass for the lighter block must exactly offset to produce equal KE to the heavier block. Finally, distractor choice D improperly applies a version of the work-energy theorem that work done by *non-conservative* forces equals the change in total *mechanical* energy (ME). While it is true that negligible friction would do negligible non-conservative work, resulting in no effect on ME, the *blower* does nonconservative work that increases both blocks' ME equally (which in this case is all KE). Q1 choice D is the only one the teachers do not discuss in this case study.

Code-line 1 (Figure 1) provides a coarse overview of the discussion. In the first half of the episode, teachers were coded as showing *discomfort and vulnerability* (code 10) in answering the question. They also *critique the prompt* (code 9) and *deemphasize the correct answer* (code 7) throughout. In the second half of the episode, they turn their discussion toward *pedagogy* (code 5), critiquing the prompt from a teaching perspective. The teachers do not discuss *physics content* (code 4) in this episode, thus avoiding the main purpose of the activity. While a key intention of the PD activity was to support physics teachers' interactions around physics content reasoning, we saw Lisa and Jessica engage in a lengthy ramp-up discussion of the merits of the question, which we believe contributed to the teachers' comfort in approaching the content-specific aspects of the task in subsequent turns. Next, we present a more detailed analysis of Lisa and Jessica's discussion through the transcript.

### 4.1.1 | Part 1: Surfacing risk and challenging the epistemic activity

To begin, Lisa and Jessica silently read the prompt for 3 min, then Jessica initiates conversation by asking if Lisa is ready to share. After a brief exchange, they share their first reactions to the prompt:

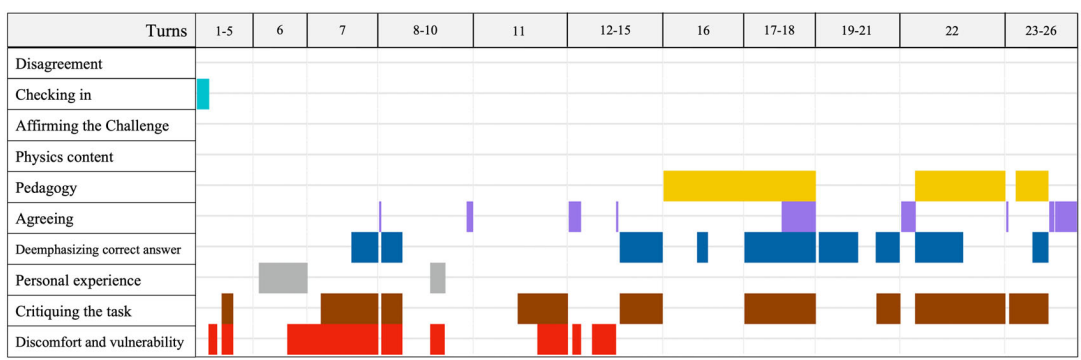

**FIGURE 1**  Code-line for Section 1 of teacher interaction. The metric for length on the x-axis is the number of characters in the transcribed spoken segment. Our protocol allowed for parts of turns to be coded. Due to some turns being much shorter than others, multiple short turns may be contained in a single heading label.

6. *Jessica*: Yeah (1.6-s pause) I mean, honestly I- when I was at [undergraduate institution] we worked with problems like this a lot. I remember always having to like read them like three times. (Eyes wide; leans in and narrows eyes to reread question.)

7. *Lisa*: I don't know, I- when I read-, I will say when I read the first *question*, I don't (frowns during a 2.6-s pause) I don't know what they mean necessarily by (1.1-s pause) *which is* (eyebrows pull together during a 1.4-s pause I don't- I don't like (shakes head) when they say these most correct and most com- complete because

8. *Jessica*: Yeah (laughs, exhales loudly)

9. *Lisa*: I don't- I don't know, I- I- I- uh I'm not su- it depends on what you're *looking* at to me but um (frowns and squints during a 3.7-s pause)

10. *Jessica*: Um so they- (1.6-s pause) they just wanna know if their b- because statement is (2.6-s pause) fully correct (raises eyebrows), which always gets me wrong cause I'm like well, maybe they (1.0-s pause) mean that, they just used the wrong words but you're right (raises eyebrows and smiles)

Here, Jessica recalls difficulties she experienced with similar problems in the past, saying "I remember always having to read them like three times" (turn 6) and "which always gets me wrong" (turn 10). This reveals a socio-emotional risk for Jessica: getting a question wrong due to tricky wording. Jessica's sense of risk could be enhanced by answering in front of a more seasoned teacher, a situation wherein the possibility for public embarrassment may feel greater.

Lisa starts by critiquing the epistemic nature of the question, disliking the definitive language of "most correct and complete" and favoring a more subjective criterion ("it depends on what you're *looking* at to me," turn 9). Both teachers narrow their eyes at the screen and lean in. Jessica additionally cocks her head to the side while reading. Based on these body language cues and intonation, we read Jessica as particularly uncomfortable in this initial exchange.

### 4.1.2 | Part 2: Mutual sharing of vulnerability and critiquing the problem as teacher peers

From Jessica's initial sharing of vulnerability and Lisa's initial critique of the question, the conversation continues on to reframe the first question as a "gotcha"-style question:

11. *Lisa*: [...] To *me* I- I feel on *edge* when I read this first question, like I feel like (1.6-s pause) like it's a gotcha (tilts head, raises eyebrows) and

12. *Jessica*: Oh yeah—I hate those questions (smiles).

13. *Lisa*: And I'm like I'm it's (gaze shifts to the side) like I'm questioning my ability to teach physics right now (gazes up, shaking head). (nods, 1.6-s pause) Ya know?

14. *Jessica*: Yeah. (Exhales loudly)

15. *Lisa*: That's when I read this I go, *Oh* (eyes wide) so there's some *right* answer and I don't know so that that's why I'm not big on *this* right now.

Both teachers' body language is loose and relaxed here. Lisa's inflection rises at the end of each turn. In her second turn, Lisa has her head propped on her hand and smiles as her eyes wander away from the screen as she speaks. Jessica smiles and laughs throughout the exchange. We read these nonverbal cues as signs that the teachers are feeling more comfortable and relaxed.

We identify three early comfort-building moves here. First, matching Jessica, Lisa shares a sense of discomfort with this question in stating, "I feel on *edge* when I read this first question" (turn 11) and "I'm questioning my ability to teach physics right now" (turn 13). Here, Lisa's display of discomfort demonstrates to Jessica that, even after 25 years of teaching, learning is an ongoing process and not knowing the answer is acceptable. The mutual sharing of vulnerability here may potentially contribute to a safe space where teachers can openly share their ideas. Second, Lisa connects her discomfort to her epistemic critique of the question, labeling it a "gotcha" question with "some right answer" and stating that "that's why I'm not big on *this* right now" (turn 15). Jessica echoes this "gotcha" critique. Questioning the epistemic activity of finding the correct answer can be comfort-building by defusing the risks of choosing the wrong answer. Third, Lisa starts to turn the conversation toward teaching, framing the task as choosing the correct answer but figuring out whether the question would encourage good discussion. This third move is yet another way Lisa challenges the epistemic activity; she invites a conversation about classroom use, approaching the problem as a teacher, which allows the pair to engage in a constructive conversation about the prompt without discussing its underlying physics content.

### 4.1.3 | Part 3: Gaining distance from risk by re-constructing the prompt as teacher peers

Jessica picks up on Lisa's challenge of the epistemic activity and the conversation turns toward the use of this question in teaching. This leads to Lisa and Jessica co-constructing more instructionally-comfortable alternatives for the prompt:

16. *Jessica*: I- I feel like I do I can see (looks up, tilts head side to side) like if they were in like a group setting to be able to just like (raises eyebrows) go through each and discuss. (2.1-s pause. Inhales, looks up, and tilts head.) It's just harder (2.3-s pause. Exhales and chews lip) when (2.9-s pause) they get something wrong how to like talk to them and be like, (1.2-s pause) you know, how to be like, "well, why'd you think that?" or something because they're already

17. *Lisa*: You've gotta word it as like "what are the, what are the, the strengths and weaknesses of each of these, um, responses?"

18. *Jessica*: Oh, yeah, so it's not like one of those true or false questions where it's like really tricky (moves head side to side)

19. *Lisa*: Yeah

20. *Jessica*: like one word is wrong and it's like, that should have been true, but that one word was different (laughs).

21. *Lisa*: Yeah, I mean, I just feel like it's (1.1-s pause) I don't know. They all have (1.1-s pause) some (2.0-s pause) part of, of the correct idea in them (pitch goes up).

22. *Jessica*: and I really like what you just said, like instead of saying which one's the best and most complete (slight head shake, eyes widen), instead, how about you give me reasons why this is incomplete. How about you give me reasons where something's wrong in this, or, if you don't find anything wrong, let me know why.

23. *Lisa*: Yeah (slight nod)

24. *Jessica*: And then that way it gives them even more discussion, cause now they have to figure out these answers too (said quickly).

25.  *Lisa*: Right, right

26.  *Jessica*: Yeah. I like that! I do like that as a way to solve that. Because, yeah (laughs).

Although Jessica begins the interaction by noting discomfort with the questions (turn 6), the conversation in the above-cited transcript seems more comfortable. What contributes to this turn? We attend to three elements that reiterate our focal moves for this part of the teachers' discussion.

First, Lisa has already shared a sense of discomfort with the question. We propose that the sharing of mutual discomfort may help the teachers feel safe to expose their vulnerabilities and take risks. Second, Lisa's epistemic critique and revision of the question alleviates Jessica's stated discomfort as a problem-solver. Jessica points out that Lisa's revised question alleviates the risk that the details in wording can make a mostly correct answer wrong (turns 18 and 20), the same risk that Jessica reported earlier from the perspective of a problem solver (turn 10). Third, Lisa's epistemic critique also repositions Lisa and Jessica as teachers discussing the pedagogical value of the question, which Jessica picks up here (turn 16). Lisa revises the question in line with her epistemic critique (turn 17), and Jessica, in turn, riffs on this revised question (turn 22). The pedagogical discussion positions Lisa and Jessica as teachers, a position that is perhaps more distant from the risks that they may experience as problem solvers.

In addition to building comfort to support a safe space for taking risks, their reframing of the question also enhances opportunities to practice and develop CKT in another way. Focusing on the strengths and weaknesses of each student explanation in the different multiple-choice options mirrors the pedagogical task of evaluating student thinking and requires teachers to practice unpacking the underlying physical ideas.

As Jessica excitedly discusses a new way to pose the problem (turns 22, 24, and 26), Lisa uses few words to respond affirmatively to Jessica's points (turns 23 and 25), leaving significant space for Jessica to complete her thoughts, while showing Jessica that she is valuing her ideas.

## 4.1.4  |  Coordinating evidence from teacher interviews: Sharing vulnerability and challenging the epistemic activity as comfort-building moves

Both stimulated-recall interviews support the assertion that risk and vulnerability underlie Lisa and Jessica's previous discussion. Lisa recognizes the risk that young teachers can feel in being accountable for knowing difficult physics content they have not mastered. Jessica states this risk in terms of "looking stupid" in front of her colleagues, but also shares how Lisa's comments helped normalize her feelings and increased her comfort within the PD program.

In her interview, Jessica shares the fear of "looking stupid" she experienced at the start of the activity with Lisa, then reflects on how Lisa's expressions of vulnerability helped her to feel more comfortable. Asked directly about Lisa's admission that she was "questioning [her] ability to teach physics right now" (turn 13), Jessica expressed appreciation and surprise for the sentiment that a more experienced teacher is still doubting her teaching abilities:

> I mean I always question my ability to teach physics, so that's definitely not anything new to me. I feel like almost every teacher should question their ability 'cause that's what helps us grow. [....] I was more surprised in the fact that [...] she has more background in teaching than I do, and I was like, if you don't know, I don't know (laughs).

 

Jessica admits here that Lisa's admission of self-doubt and vulnerability normalized those sentiments for Jessica, who respected Lisa's extensive experience as a teacher.

Lisa was aware of Jessica's vulnerability and the pressure teachers feel to get the right answer. After viewing a video clip from section 1, Lisa shared:

> Well, it reminded me, as I'm listening to Jessica talk, and my kind of hesitancy, I think [...] whether we are the teacher or the student or whatever, is that we want to get the right answer...I think new teachers feel bad when they don't remember every little thing. But it's not a bad thing, and in fact, I have spent more time thinking about how to teach energy than any other topic in physics. I think it's the hardest, it's really easy to learn how to solve energy questions without really understanding energy. And I think a lot of us come out of high school or college or whatever thinking we really understand energy, and then, and then when we have to explain something the truth comes out.

Lisa and Jessica's interviews also revealed that their critique of the epistemic activity—a comfort-building move in the teachers' discussion—was connected to valuing students' ideas as a teacher. In her interview, Lisa critiqued the binary classification of best and worst explanation. Instead, she connected the discussion of strengths and weaknesses to valuing students' ideas.

> *Lisa*: When a kid gives you 'this is what I think,' the hard part as the teacher is how do you respond to that? Do you, do you say, 'no, you're wrong, and this is how you should be thinking?' or, obviously not, what sorts of questions, how can you guide a student into critiquing their own thinking and coming up with something to follow up [....] How can you give kids the freedom to explore what they're thinking but make sure that it's something worthwhile for them to do?

Here, Lisa casts valuing students' different reasoning as a rationale for critiquing the epistemic activity of choosing the best answer. In addition, Lisa explains that she was initially struck by the lesson design in the prompt as opposed to the physics content, and, in the interview, uses this approach to reflect on her own teaching. This process of self-reflection continues as Lisa describes her initial thoughts on the problem:

> *Lisa*: I really liked how this teacher [in the prompt] was asking her students to put into words what they were thinking, [...] I don't think I ask my students to verbalize their thoughts enough [...] to make a claim and justify it. I think I could do better with that.

Jessica echoes this in her reflections on an exchange from the end of the episode, where she confidently led a discussion about a problem and smoothly blended physics concepts with pedagogy in her remarks. Similarly, Jessica reported liking the pedagogical approach of asking students to build on their thinking, saying: "I think I was very excited about [....] letting them have a wrong answer and then having the whole class come together with the right answer."

Jessica confirmed in her interview that critiquing the task increased her confidence that she could participate in and contribute to the discussion. She explicitly noted her appreciation of Lisa's suggestion as a comfort-building move:

*Jessica*: I remember thinking that as a student, how would I feel answering those questions, and then when she rephrased that question and was like "tell me why you think things are wrong" or something, I remember being like, oh, yeah, as a student, I would probably enjoy that more (laughs) than being like, this is right, and then being all insecure about my correct feelings rather than saying I know why this one's wrong.

Lisa does not explicitly connect her critique of the epistemic activity to comfort building, either for her students or Jessica during this episode. It seems that comfort-building was the outcome of Lisa's epistemic critique, but not an explicit motivation for it.

## 4.2 | Episode section 2—Collaboration-centered problem-solving and storytelling

Code-line 2 (Figure 2) provides an overview of section 2 of teachers' discussion of Two Blocks Q1 (turns 27–46). Here, they discuss *physics content* (code 4) for the first time. Through the transcript, we will show that Lisa initiates collaborative problem solving by starting to think aloud about the content of the Q1 multiple-choice options and intermittently turns to solicit feedback from Jessica. Although this starts the discussion of physics content, *discomfort and vulnerability* (code 10) still arise in this episode in between the content discussions. After the first appearance of *discomfort and vulnerability*, the teachers share stories of their *personal experience* (code 8) between bouts of physics content discussion. We will show that this storytelling is a comfort-building move that supports safety in the face of Jessica's discomfort.

### 4.2.1 | Part 1: Discomfort calls off physics content discussion

Following section 1, Lisa pivots toward the physics content of the multiple-choice options to the question for the first time, opening up the risks of being accountable to knowing physics content for the teachers:

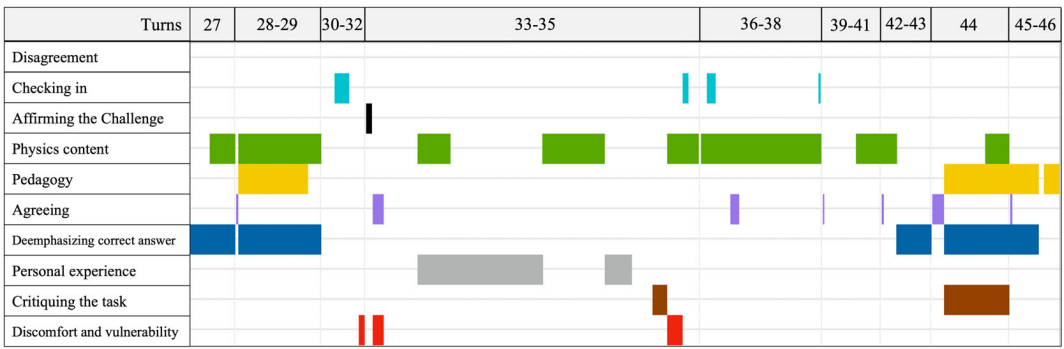

**FIGURE 2** Code-line for Section 2 of teacher interaction. The metric for length on the x-axis is the number of characters in the transcribed spoken segment. Our protocol allowed for parts of turns to be coded. Due to some turns being much shorter than others, multiple short turns may be contained in a single heading label.

27. *Lisa*: Well, I mean, cause when I'm looking at this, (1.1-s pause) I mean, I guess I would say the first one is *not* a very good answer, because it's focusing on *time*,

28. *Jessica*: Yeah.

29. *Lisa*: and energy isn't really about time. So the kid is understanding though (leans back slightly and lifts hands to front on each side, palms up, hands slightly curved) that it's equal amounts of *energy*, they've got that *idea*, but they're just not (shakes head, 1.8-s pause), you know, they're focusing on the wrong, (1.9-s pause) the wrong, uh, increment, interval I guess. Where I guess C is better in that regard.

30. *Jessica*: Yeah.

31. *Lisa*: 'Cause it's talking about position. (4.3-s pause) Right? Because it would be like a work idea? (Moves head side to side)

32. *Jessica*: Oh (shaky), yeah, okay that makes sense, sorry (smiles), I have to (tilts head)

33. *Lisa*: No! No, I get it. (shakes head slightly)

Lisa starts by (correctly) discussing the physics of the answers. She names the core ideas in the work-energy theorem, distinguishing why choice C is correct and choice A is incorrect. Specifically, changes in energy correspond to the amount of work done—the product of force and distance, not force and time.

Here, Jessica evidences the presence of a risk, responding only "yeah" to Lisa's reasoning in turns 28 and 30. After a 4-s pause where Jessica does not respond, Lisa phrases a more direct question ("Right?") in turn 31, followed by connecting her reasoning to the physical concept of work. Through Lisa's intonation and body language, she invites Jessica to respond. Jessica then apologizes for presumably needing time to consider Lisa's reasoning (turn 32). During this turn, Jessica hunches over and nervously smiles, perhaps to create time to process, or perhaps to defer the lead of conversations about content to Lisa. Jessica's lack of response and nervous apology signals the potential re-emergence of risks and, in turn 33, Lisa switches from discussing content to a repair move.

## 4.2.2 | Part 2: Sharing personal stories smooths tension by normalizing confusion

Following Jessica's expression of discomfort and apology, Lisa responds by normalizing and relating to Jessica's confusion with a personal story about her own learning process:

35. *Lisa*: So right there, actually, our conversation right there is exactly what you'd want to be talking about. That idea of energy, energy and time is momentum and impulse, energy and position is work and energy. That idea (shakes head), I didn't have straight in my head (shakes head slightly), or that clear of it being straight when I first started teaching physics. And then as I worked with kids who had trouble figuring out, "Oh, do I use energy or do I use momentum in this situation?" Then I figured out (leans back slightly, 1.7-s pause) how to explain to them what you use. I'm like, if what you're looking at is *time* changes, then you should focus on impulse and momentum, but if you're looking at *position* changes, functions of position, then that leads you to ideas of energy. But, so, I wouldn't have been thinking about that (leans in) when I first started teaching physics. So that's why this *is* a good opportunity for discussion (pauses for 2.0 s, then shakes head as she continues) but again I just don't like the wording of the question. (1.5-s pause) I don't know about B and D. Which one, I don't know, what do you think? Which one do you think would be...

After directly reassuring Jessica ("...our conversation right there is exactly what you'd want to be talking about..."), Lisa shifts to a personal anecdote about gaining competency with the material, highlighting that she had not mastered these ideas when she started teaching physics.

Embedded in this comfort-building anecdote is a further elaboration of the physics content: *time* being related to impulse and momentum while *position* changes are related to work and energy, and the explicit recognition that this question is a good opportunity for discussing and differentiating these ideas. After again critiquing the wording of the question, Lisa continues deliberating ("I don't know about B and D") and then asks for Jessica's opinion, signaling that she values and respects Jessica's ideas. These moves may function to normalize being challenged by the question as a new teacher, explain concepts relevant to the problem, and open space for Jessica to share her thinking. In her interview, Jessica revealed she gained confidence in learning that Lisa also had not mastered the nuances of energy and momentum problems as a new teacher: "She helped me have confidence when she said, 'oh, I felt the same way when I first started. I didn't know either.'"

### 4.2.3 | Part 3: Thinking aloud opens space for content reasoning

In response to Lisa's solicitation of Jessica's ideas, Jessica begins thinking aloud about choice B. In what follows, the pair eventually establish (correctly) that choice C is the best answer, while also noting the merits of choice B. Although B has aspects of correct understanding, the teachers correctly see that, while differences in final speed could account for the differently massed blocks having equal energies, this is the *result* not the *cause*—thus it cannot be considered most correct and complete.

38. *Jessica*: Well, okay, so if you're thinking about (1.2-s pause) the different masses, then you have to think about the speeds, (1.2-s pause) so I actually kinda like B (says "B" quietly). "The final speed of the lighter block compensates for its smaller (eyes narrow) mass." (3.4-s pause) So (2.6-s pause) I mean, if they have the same KE, I don't think that (1.8-s pause) that explains if they have the same KE, that just explains what the smaller block did (1.9-s pause), right?

39. *Lisa*: Yeah

40. *Jessica*: Okay, cool.

41. *Lisa*: I mean they are, it is an important *idea*, though (lifts eyebrows), of this discussion, you know, like, (1.6-s pause, purses lips) how *is* it possible (nods slightly) that these two masses (1.8-s pause) *can have* the same amount of energy?

42. *Jessica*: Yeah

43. *Lisa*: To me it's not relating it to the force (nods head slightly). But it's still a good i- (shakes head slightly) you know, they're (moves hands in circles, palms up, laughs), it's like the kind of thing I ha- I just (leans in), yeah, I suppose that would be it.

44. *Jessica*: (smiling) I *really* like what you said though. Is instead of having them find the bes- (lifts eyebrows), the most correct or something, instead having them (lifts chin, lowers gaze) give an explanation of why (1.1-s pause) this statement is telling you (slight nod) about what, like, tell me (tilts head, lifts gaze) exactly what this statement tells *you* about the blocks and the KE. (moves head side to side slightly)

45. *Lisa*: Yeah

This is the first time that Jessica reasons through the physics content with Lisa in detail. However, Jessica's language and behavior in turn 38 may indicate she has shared before feeling confident in her reasoning. First, her language distances her from her answer choice ("I actually kinda like B", turn 38). Then, after re-reading the problem, Jessica asks for Lisa's confirmation (end of turn 38: "…right?") when pointing out a weakness of choice B. Lisa validates Jessica's concerns about B while continuing to value aspects of the incorrect student responses (turn 41), leading Jessica to once again position herself as a teacher in critiquing the epistemic activity.

Together, the pair unpacks the issues in choice B related to CKT-E. They agree that B is an explanation for how these two blocks with different masses *can* have the same KE (turns 38–40), but that it does not give a principled reason why the kinetic energies *must* be equal. Still, in line with the teaching task of recognizing productive ideas in student reasoning, the teachers view the post-hoc explanation here as "an important idea" (turn 41). Thus, they explain the most common reasoning error seen in Etkina and colleagues' prior deployment of this question, while also identifying the underlying, productive elements of physics reasoning.

To summarize this section, Lisa initiates collaborative problem solving by thinking aloud about the answer choices and encourages Jessica's participation explicitly (by asking for it), and tacitly (by modeling that it is okay to show uncertainty in one's reasoning). When Jessica's statements show risk and vulnerability, Lisa engages in storytelling—sharing her own history with teaching the topic and normalizing confusion about physics content early in teaching, while also explaining relevant physics content. We interpret thinking aloud and storytelling as a comfort-building moves. Subsequently, when invited by Lisa, Jessica leads discussion of the physics content for the first time, and her reasoning is affirmed by Lisa. By the end, the teachers have discussed CKT-E by evaluating three out of the four student explanations. They identified the errors and productive ideas in incorrect choices A (which failed to differentiate the impulse-momentum theorem from the work-energy theorem) and B (which used compensatory reasoning to describe how energy could be conserved but did not explain why energy must be conserved), and articulated the work-energy theorem as the relevant physical principle best embodied in correct choice C. Discussion of the distractor choices allowed the teachers to practice and develop their CKT by identifying the productive seeds of correct physics thinking which can be built upon in these responses, and evaluating what could be corrected and refined in the underlying student reasoning.

Until this point, Lisa has guided the conversation. Although there have been significant contributions from Jessica throughout the episode, exchanges about content follow a very specific pattern: when Jessica talks about content, it is in response to a prompt from Lisa. In the following episode section, in which the teachers discuss Two Blocks Q2, this dynamic shifts.

## 4.3 | Episode section 3—Cycle of disagreement and repair

The code-line for section 3 (Figure 3) gives an overview of teachers' discussion of Two Blocks Q2 (Table 2) in turns 56–119. *Physics content* (code 4) is sustained at the beginning and end of the section as teachers discuss Q2. The small spaces between the bars in code 4 represent changes in speaker, showing the discussion of content is more balanced between the two teachers. *Disagreement* (code 1) surfaces for the first time (turns 60 and 114). Similar to section 2, *discomfort and vulnerability* (code 10) are present and are present at transitions between discussing *physics content* and sharing stories of *personal experience* (code 8).

Q2 is designed to be more pedagogically complicated than Q1, as it requires evaluating which proposed laboratory scenario will leave both blocks with the same final KE. The correct choice is D, since the same compressed spring will do the same amount of work (exerting the same average force over the same distance) on each block as it decompresses. Alternatively, because spring forces are conservative, one could reason that the potential energy (PE) stored in the compressed spring is transferred into the KE of the block when the spring decompresses. The incorrect choices each represent a scenario where the hypothetical student's explanation would require revisions. Correcting the reasoning in distractor A would require the student to differentiate the same force applied over the same distance (equal work done, resulting in equal KEs) from the same force applied over the same time (equal impulse imparted). Distractor B would require the student to reason that the heavier block experiences greater frictional force, and thus more work done by friction, so the final KEs would not be equal. Finally, in distractor C, the student must recognize that different changes in gravitational PE will, all else being equal, result in unequal final KEs. In discussing all four choices, Lisa and Jessica practice and develop the CKT-E required to guide students in correcting and refining initial, incorrect thinking.

In this section, it is notable that Jessica starts the physics content discussion, a shift from the previous two sections. Here, we see three different results when one teacher shows discomfort and vulnerability: (a) Lisa opens by showing discomfort, opening the door for Jessica to lead into the problem; (b) As before, Jessica expresses discomfort with the physics content, calling off that topic; and (c) Lisa begins to take Jessica's expressions of vulnerability as signals she is ready to discuss physics content.

### 4.3.1 | Part 1: Disagreements surface, followed by a lengthy repair

As the teachers start reading Two Blocks Q2, Lisa states that "this one even confused me even more." After they quietly read the problem for 22 more seconds, Jessica offers an initial idea for the problem:

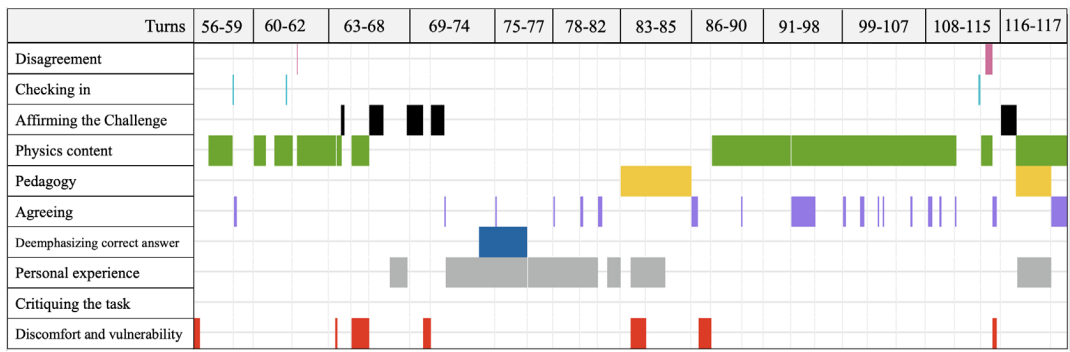

**FIGURE 3** Code-line for Section 3 of teacher interaction. The metric for length on the x-axis is the number of characters in the transcribed spoken segment. Our protocol allowed for parts of turns to be coded. Due to some turns being much shorter than others, multiple short turns may be contained in a single heading label.

57. *Jessica*: So if they were pushing hard (scratches head), that means they have the same force (1.1-s pause, brow furrowed, eyes narrow) being pushed on them, which doesn't necessarily mean they have the same kinetic energy, right?

58. *Lisa*: Right. (nods slightly)

59. *Jessica*: Okay, cool. (softly reads to herself with narrowed eyes for 8.7 s)

60. *Lisa*: Although I see if they're pushed for the same amount of distance (hand at side of face, chin lifted, gaze lowered)

61. *Jessica*: If it said that, yeah (brow furrowed, softly reads to herself) *but* (tilts head, eyebrows lifted) same amount of distance doesn't (shakes head), 'cause they're different si-, masses. So you're doing more work. (tilts head side to side with furrowed brow)

62. *Lisa*: It'd be the same ener-, no, it'd be the same work. So if you have, if the blower pushes equally on the blocks, (1.2-s pause) then the same, force is the same (small, single nod), and if it's the same finish line (small, single nod), it's the same distance (small, single nod). So they would have the same *work* (small, single nod).

63. *Jessica*: Force times distance, yeah (frowns with lifted gaze).

64. *Lisa*: And then

65. *Jessica*: Sorry! (Laughs)

66. *Lisa*: for the kinetic energy. *Right!* (lifts eyebrows) It's tricky! (Shakes head) So the kinetic energy would be the same.

Here, the teachers come to restate the underlying ideas in the work-energy theorem without referring to any specific answer choice. Jessica is taking a think-aloud problem-solving approach that Lisa modeled before; she verbally offers some initial ideas that represent the start of productive reasoning, correctly recognizing that equal applied force does not imply equal (final) kinetic energies. When Lisa initially agrees with her (turn 58), Jessica seems relieved and immediately goes back to reading the problem (turn 59), making space for a turn by Lisa. As Lisa starts to elaborate on Jessica's initial idea (turn 60), Jessica continues to share her ideas and check that she is on the right track (turn 61). This stands in contrast to previous sections where Jessica was more tentative in offering her ideas, only doing so when explicitly invited to by Lisa.

In turns 61–66, the pair again come to the work-energy theorem, correcting an incorrect idea suggested by Jessica that unequal block masses indicate unequal work done, even with the same force exerted over the same distance. Though incorrect, the value of voicing this incorrect idea is that it creates an opportunity to practice and develop their physics content knowledge through subsequent corrective discussion. It appears that Jessica was not considering the application of the work-energy theorem in her initial reasoning. Though Jessica comes to realize her error (turn 63), her apology (turn 65) indicates her discomfort in being incorrect; Lisa again follows this sign of vulnerability with a comfort-building repair ("Right! It's tricky...", turn 66).

Comfort-building moves continue as the teachers tell stories about their experiences as physics students and teachers:

67. *Jessica*: (exhales) I just feel like I've never thought of kinetic energy, (1.2-s pause) I'm not, it's like hard for me to think of kinetic energy right now in that way.

68. *Lisa*: This is actually to me, energy is the hardest subject to teach in physics. It's interesting that they picked this. That's what I found. The way that I teach energy has changed *more* than any other topic in physics, so it's not an easy (shakes head)

69. *Jessica*: (laughs)

70. *Lisa*: you can't get your fingers on it, you know? It's a hard, it's a hard topic.

71. *Jessica*: I feel like it's been a while since I've (nervous smile)

72. *Lisa*: Yeah, no, absolutely, and since you've thought about it at this level, too!

73. *Jessica*: (laughing) Yeah!

74. *Lisa*: That's, and the thing is that in college—this is what I found when I started teaching—you know, in college it wasn't this kind of stuff. It was all about getting the right answer and getting done. You know, like, you weren't thinking this deeply about questions. Like when you were taking your physics class (laughs)

75. *Jessica*: Yeah

76. *Lisa*: like your intro to physics classes on this kind of stuff, you were a student, I was a student, just trying to get through it. And so I didn't think this deeply about any of this stuff (shaking head). Yeah.

The apologies that Jessica gives when taking time to think, or giving the wrong answer, likely arise from the expectation that physics teachers should be able to quickly and correctly answer these questions. Lisa deflates this tacit expectation by (a) stating that energy is a difficult topic (turn 68), (b) stating that she did not think deeply about "this kind of stuff" in college (turn 76), and (c) describing how her teaching experience has changed how she teaches energy (turn 68).

In turns 77–86, which we summarize here, Jessica relates to Lisa's college experience and shares her own experience of not knowing the answers to challenging physics questions in college. The teachers jointly acknowledge the experience of not understanding physics deeply as a student, while naming the shared value of challenging students to deeply understand material through activities like discussing the Two Blocks questions (turn 85—Jessica: "...I've seen the rewards from students doing stuff like this and I'm like, 'that's amazing!'").

## 4.3.2 | Part 2: Working through tension and discomfort leads to a fluid exchange and an "a-ha" moment

After this storytelling exchange, the pair start reasoning about the choices in Two Blocks Q2. The exchange is fluid with equal contributions from both teachers:

87. *Jessica*: (reads quietly to herself) I feel like I just need to like (leans in, chin down, rests face on hand, 2.0-s pause) be learning the subject right now to actually (laughs)

88. *Lisa*: So *I'm* thinking, I'm thinking (raises eyebrows) that the last one is the *only one*.

89. *Jessica*: What you said about, okay, so work equals...force times distance, *but* (1.1-s pause) the only thing about that is (1.4-s pause) the kinetic energy. I mean there's not other energy going on here, so, yeah (shakes head).

90. *Lisa*: So but in the, (chin up, gaze down, 1.3-s pause) let's see, so in the first situation, it's talking about time (shakes head), again, that's not what we're talking about.

91. *Jessica*: I mean, there is-

92. *Lisa*: It's gonna take *longer*,

93. *Jessica*: but the

94. *Lisa*: but the *motion* is gonna take longer for (1.9-s pause) the bigger block. (3.3-s pause) Like, the little block is gonna win the race.

95. *Jessica*: *Oh!* So now I understand what you meant about the whole (looks down)

96. *Lisa*: Right. This isn't about-
97. *Jessica*: thinking about the same force and the same distance.
98. *Lisa*: This is that, when they reach the end, (1.0-s pause) they have the same amount of kinetic energy, (1.6-s pause) whether they're a big block or a small block. (2.9-s pause) But the big block is going to lose the race.

In the turns above, Lisa correctly states that the last multiple-choice option, D, is the only correct choice but does not explain why (turn 88).

Here, the teachers continue to construct refinements of the reasoning errors embedded in the incorrect choices. In turn 90, Lisa turns the conversation to choice A, again pointing out that time is not the relevant issue for energy. This idea is key to differentiating work-energy ideas from impulse-momentum ones. In turns 92–98, Lisa elaborates on this differentiation, correctly explaining that both blocks can gain the same KE but take different amounts of time to travel the same distance.

In this part, Jessica continues to get more engaged in the physics content of the question and responses. In turn 89, Jessica takes up Lisa's point (all the way from turn 60) about work depending on force and distance, and applies it to this situation. During this discussion differentiating distance and time, Jessica expresses an "a-ha moment" (turns 95 and 97), now understanding Lisa's previous explanations regarding force and distance. The pair next evaluate the incorrect answer choice B, and Jessica's engagement with Lisa's ideas deepens:

99. *Jessica*: Yeah, okay. (Nodding, 1.4-s pause) Okay. Um, so, constant force for the same time interval. I, yeah, you're right, so I don't think A would be right. B would be, (1.1-s pause) like you were saying (quietly reads to herself)-
100. *Lisa*: (head slightly tilted) So this one, though, the frictional force *is* gonna be different
101. *Jessica*: Yeah.
102. *Lisa*: for the two blocks,
103. *Jessica*: Yeah (nods).
104. *Lisa*: because they are different masses, so, so that's (shakes head), you know, that's not gonna-
105. *Jessica*: That-
106. *Lisa*: The big one's gonna lose more energy (1.1-s pause) than (1.5-s pause) than the smaller one (puts hand down).
107. *Jessica*: Oh, okay, yes, you're right (looks down and to the right). And so they, it won't get up to the same amount of kinetic energy. But it will have the same amount of work
108. *Lisa*: Right, they'll have the same amount of work (nodding) but
109. *Jessica*: Now the energy's broken up into two parts.
110. *Lisa*: Right.

In terms of discomfort and vulnerability, this part contrasts with previous ones where Jessica apologizes for not understanding or thinks silently to herself. Though Lisa again leads the discussion of physics content, Jessica engages with and expands on Lisa's statements. In turn 99, Jessica evaluates Lisa's argument against choice A and expresses agreement. In turn 107, Jessica builds on Lisa's argument, and Lisa and Jessica come to agree in turns 108–110. The continued comfort-building moves throughout the episode so far support Jessica's content reasoning seen here by addressing the socio-emotional risks of engagement with challenging physics content.

Though the teachers correctly conclude that choice B is incorrect, their reasoning is not completely correct. Lisa gives a correct explanation: the magnitude of the frictional force is mass dependent, so the two blocks will experience different amounts of frictional force. Although both blocks will experience a frictional force that does negative work contributing a negative change in KE, the different frictional forces will do different amounts of work causing different changes in KE (turn 106: "The big one's gonna lose more energy..."). Jessica synthesizes Lisa's ideas, agreeing that she is correct and that the scenario in choice B will not lead to equal final kinetic energies for the two blocks (turn 107). Yet, in turns 107–108, the teachers also state an unrefined idea—that both blocks experience the same amount of work. In actuality, both blocks experience work from two sources: work done by the blower and work done by friction. It is true that the work *done by the blower* on each block is the same, but the work *done by friction* on each block is different, so the net work done on each block is different. Though Jessica and Lisa make a correct argument for rejecting choice B, there is more refinement needed in differentiating the work done by the blower force and the work done by friction. In turn 109, Jessica hints that the seeds of splitting these two contributions are already planted by saying "now the energy's broken up into two parts."

### 4.3.3 | Part 3: Playfulness and affirming the challenge mitigate the risk of being incorrect

In this final part, Jessica offers an approach for choice C that Lisa corrects, which is again followed by a sign of vulnerability by Jessica and a comfort-building repair by Lisa, this time with a playful tone:

113. *Jessica*: (quietly reads to herself, grimaces) Okay (said slowly), so *now* on C you have to do like a force body diagram (raises eyebrows) to break this one up and then see if the energies are getting lost in different forces again, right? Or-
114. *Lisa*: You *can*, but actually, I would just think of it in terms of potential energy (gaze up, shaking head slightly), so
115. *Jessica*: Oh, yeah. Damn it! You're really good at this (smiles).
116. *Lisa*: I'm saying, you're gonna get this Jessica (both laugh) you're out of your (snaps fingers), you're out of your zone here, but no, but see this is what students do, is they start thinking, oh it's an incline question so I gotta start thinking about force components and all that rather than recognizing, *nope!* (shakes head) We are talking about energy.
117. *Jessica*: (smiling) And it's so much easier, always, to go the energy route! 'Cause you don't have to do much.
118. *Lisa*: Yeah, I'm pretty sure it's just D, because if you got a compressed spring, there's gonna be a certain amount of force ... average force I guess, that's released as it moves... so it will leave, no matter what it's gonna leave the spring with the amount of kinetic energy equal to the energy stored in the spring originally.
119. *Jessica*: Yeah, yeah, I believe that.
120. *Lisa*: But again, I think they're good questions! Like I said, to me, we just had a very good discussion about it.

The discussion continues to be productive for practicing and developing CKT-E. In considering choice C, the teachers raise the idea of using potential energy rather than work done for the first time (turns 114 and 116). For conservative forces like gravity and spring forces, work done and change in potential energy are two conceptually different, but mathematically equivalent, methods for considering effects on KE. This connection is an important piece of understanding work and energy. In turn 118, Lisa also links these two methods when correctly stating that choice D is the only correct choice (work method: "...there's gonna be a certain amount of... average force...as it moves"; potential energy method: "...it's gonna leave the spring with the amount of kinetic energy equal to the energy stored in the spring originally."). The episode closes with Jessica agreeing with this reasoning (turn 119). Here, Jessica's growing engagement continues to contribute to the group's engagement with physics content. Jessica's proposal to conduct a force analysis creates a context for the teachers to compare and connect force analysis against energy analysis.

In contrast to their previous discomfort in this episode, the teachers take a playful, supportive tone in their discussion. Both teachers smile and laugh in the discussion (turns 115–117). Potentially threatening statements like "you're out of your zone here" (turn 116) do not appear to raise socio-emotional risks and shut down the conversation, as Lisa's tone conveys a warmer and more understanding meaning. Aligned with this interpretation, Jessica stated in her interview that Lisa's comment did not bother her: "No, 'cause I felt that way (laughs)! I was out of my zone. Like I felt like I kept trying to use grad physics to solve this high school problem."

# 5 | DISCUSSION

Our analysis focused on how two teachers applied comfort-building moves within a risk-rich science content reasoning activity. This investigation was conducted in response to a call in the research community to examine the contextualized role of affect and how individuals generate, sustain and experience *affective phenomena* (Avraamidou, 2020; Curnow & Vea, 2020; El Halwany et al., 2021 and Lanouette, 2022). Through an exploratory case study, we proposed four comfort-building moves that can support the development of a safe space for teachers to discuss physics content and pedagogy: *Challenging the epistemic activity*, *revealing vulnerability*, *collaboration-centered problem solving*, and *storytelling*. Stimulated-recall interviews supported our interpretation that the pre-service teacher experienced initial socio-emotional risks but developed a sense of comfort over the course of the interaction. We argue that comfort-building moves supported the eventual freer, more balanced exchange of ideas between the teachers. The increased teacher discussion of the physics content created greater opportunities for *disagreement and repair* toward the end of the interaction, which allowed the teachers, especially Jessica, to practice and develop their CKT-E. More broadly, these findings suggest the usefulness of continued study into comfort-building moves for alleviating socio-emotional tensions that can otherwise obstruct teacher learning.

## 5.1 | Research questions 1 and 2: Socio-emotional risk and comfort-building moves

With respect to Research Question 1 (How can socio-emotional risks related to science content surface in secondary science teacher PD discussions?), we attribute Jessica's initial sense of risk and discomfort to (a) encountering a physics problem she initially did not know how to solve,

(b) the tacit expectation that physics teachers should quickly and accurately be able to answer such a question, and (c) her pairing with a more experienced teacher.

With respect to Research Question 2 (What conversational, comfort-building moves can teachers use to mitigate these risks and support a safe space for discussing scientific content and reasoning?) we observed two classes of moves that contributed to Jessica's eventual comfort in taking risks: (a) moves that diverted the conversation away from tense or stressful topics, or risk-reducing moves (e.g., epistemic distancing, calling off a tense debate, or saving face for oneself or another) (Conlin & Scherr, 2018; Goffman, 1955; Sohr et al., 2018), and (b) moves that conveyed and built empathy, care, and trust, or risk-embracing comfort-building moves (e.g., empathizing with another's epistemic stance or offering encouragement) (Appleby et al., 2021; Jaber et al., 2018). We observed these moves working in concert to build the discursive partners' collective tolerance of risk.

Jessica also reported feeling more comfortable once she began talking to Lisa. Throughout the episode, Lisa was the primary initiator of comfort-building moves. Lisa's early comfort-building provided an alternative to correctly answering the questions and a safer mode of engagement for both teachers. Some of Lisa's comfort-building moves—revealing vulnerability, in particular—also serve to downplay Lisa's expertise and power as a veteran teacher in the conversation. At first, Jessica's comfort did not translate directly into greater risk-taking but did allow her to engage with Lisa; she joined Lisa in questioning the epistemic activity and contributed ideas about the problem through a pedagogical lens.

As the conversation evolved, when bouts of physics content discussion reached a certain threshold of risk (punctuated by verbal expressions of discomfort), the teachers shared stories that served to normalize insecurity about physics content and knowledge. With more repetitions of this cycle, the pair's "tolerable" threshold of risk appeared to increase. They seemed more comfortable sitting in tension and disagreeing. Their content-based conversations became richer, more sustained, and their turn-taking more balanced. As indicated by her reflective interview quotes shared in the introduction, Jessica became more comfortable, starting to have her own ideas rather than letting Lisa lead the discussion, and better understood the physics content as the interview progressed. By the time the pair reached Two Blocks Q2, Jessica became a more equal participant in the physics content discussion. In this way, comfort-building opened space for the teachers to practice and develop their CKT-E in a PD setting, even in the face of initial feelings of socio-emotional risk.

Existing work at the intersection of social risk and epistemology in group learning scenarios commonly focuses on moves that *reduce* conversational tension (e.g., Andriessen et al., 2013; Appleby et al., 2021; Asterhan, 2013; Conlin & Scherr, 2018; Jaber, 2015; Sohr et al., 2018; Zhang et al., 2021). Though it is true that such tension-reducing moves can be comfort-building moves within a group, we conjecture that when these moves appear in regular combination with intentional displays of care and empathy, group members may begin to not only tolerate moments of heightened social risk, but in fact *embrace* them.

Specifically, our analysis revealed that Jessica and Lisa continued to surface insecurities about their physics content knowledge throughout the interaction even though Jessica reports, in her interview, feeling more comfortable in the conversation by the end. Early on, when Jessica surfaced these insecurities, Lisa guided the conversation away from the prompt, effectively calling off such tense moments. In these breaks from discussing content, Lisa frequently affirmed Jessica's sentiments by sharing her own vulnerable orientation toward physics concepts (epistemic empathy), exchanging personal stories with Jessica that normalized physics teacher insecurities (social caring), and reframing the conversation to focus on pedagogical fitness of the prompt (epistemic distancing for oneself and for each other). As a result, even though Jessica apologized or expressed discomfort at about the same rate and intensity as she did at the interaction's outset, over time the pair could lean into those tense moments instead of diverting the conversation away from them. This example shows how successful comfort-building can be seen as increased engagement with and reasoning about challenging problems in the face of continued signals of insecurity and risk in the group.

## 5.2 | Comfort-building opens up space for practicing and deepening content knowledge

Importantly, this case illustrates how comfort-building moves can open up space for exploration and deepening of content knowledge. Even physics teachers can have difficulties with physics content and reasoning, as has been shown for the two CKT-E items discussed here (Etkina et al., 2018). This is one reason why physics teacher PD programs can have teachers act as students in learning activities: to create opportunities for those teachers to actively develop their pedagogical content knowledge (e.g., Etkina, 2010). Here, teachers practiced and applied their CKT by evaluating hypothetical student explanations.

Although this case is one brief excerpt of Lisa's and Jessica's reasoning in the PD program, their reasoning includes articulation and differentiation of the central ideas of work and energy. Though both teachers may have benefitted from evaluating the reasoning behind the hypothesized student ideas embedded in the answer choices, Jessica may particularly benefit, as her reasoning appears less fluent and undergoes more corrections than Lisa's.

## 6 | CONCLUSION AND OPEN QUESTIONS FOR FUTURE RESEARCH

Using a rich episode of teacher collaboration, we abductively elaborated a mechanism (see Coffey, 2018 ch. 6, p. 12) for comfort-building in collaborative groups. This paper focuses on comfort-building moves that support socio-emotional *and* intellectual risk-taking, illustrating how attentiveness to emotion fostered deep intellectual and cognitive work for the teachers. This work builds on and connects to existing theoretical models of how collaborative groups identify, respond to, and mitigate risk when approaching challenging science questions, adding a case to a growing body of empirical support for these models.

Future efforts can explore questions about longitudinal effects of comfort-building across long time scales and activity contexts, which this single episode does not address—for instance, how can comfort-building interactions between Lisa and Jessica seen in this case impact the collaborative dynamics within the larger teacher PD program in which this interaction is embedded? In particular, future research can investigate how social relationships and histories between teachers can impact risk and comfort within a teacher PD program. While it remains common to construct case studies of how groups navigate socio-emotional risk within a bounded learning episode (e.g., Conlin & Scherr, 2018; Sohr et al., 2018), there is a growing focus on how the social and empathetic relationships built between group members over time can be catalysts for learning (e.g., Appleby et al., 2021; Jaber et al., 2018).

In addition, future research can investigate how differences in participants' power and prior experience can impact comfort-building dynamics. In this case study, Lisa, the more experienced teacher, is also the one who initiates the comfort-building moves, but to what degree does her higher status in the group increase access to comfort-building moves? For instance, challenging the epistemic activity is exercising power (power in the sense that one is allowed to critique and redefine an activity) and revealing vulnerability is safer from a higher status position. Could a group of more homogenously positioned novice teachers as successfully initiate and sustain comfort building in the face of socio-emotional risks? Future research can investigate how power and positioning in a group can afford and constrain access to comfort-building moves. Another open question regards the relationship between who initiates comfort-building moves and who voices discomfort and risk. It is sensible that, in this case, Lisa would make comfort-building moves in response to Jessica's signaled discomfort but could Jessica have made comfort-building moves on her own behalf? In what ways could a less experienced teacher who is feeling discomfort have initiated the comfort-building moves?

This case investigates socio-emotional risk and comfort building as related to a persistent problem of practice for in-service secondary science teacher PD programs: the challenge of interweaving content learning for teachers who are nominally already content experts (Etkina, 2010; Etkina et al., 2017, 2018). Feelings of insecurity among teachers may be further heightened when such PD programs create opportunities for new or out-of-field teachers to interact with experienced ones (see Grossman et al., 2001; Liu, 2013; Netz, 2020; Sutton & Shouse, 2019). Can comfort-building be deliberately designed for in secondary science teacher education settings? Future research can explore a several factors. First, rather than coming up with the correct response or evaluating responses based on correctness, problem tasks might be explicitly framed by PD facilitators as a discussion of pros and cons of different responses, a comfort-building framing that the teachers in our case take on themselves. Second, future research can explore potential consequences of homogenous and heterogenous grouping of teachers according to experience. This case illustrates the work of a novice-veteran pair, so investigating the degree to which these comfort-building moves translate to other combinations of teacher expertise (e.g., novice–novice, veteran–veteran, or pairs with an out-of-field teacher) is a valuable research direction. Third, future research can uncover other aspects of teacher's enacted identities beyond teaching expertise that can be consequential for how they build comfort and manage socio-emotional risks. Although there is no one-size-fits-all approach to groupings for these activities, understanding potential impacts of these groupings can inform how PD facilitators organize and facilitate teacher groups. Fourth, the impact of PD facilitator moves can be investigated in terms of how they facilitate teacher comfort and teacher–teacher comfort-building. Some candidate facilitation strategies we raise for future investigation are to: (a) structure activities to involve less socio-emotional risks for new teachers to contribute meaningfully; (b) implement activities that promote teachers' expertise (such as having teachers share a successful practice or activity from their teaching); and (c) have facilitators explicitly promote and implicitly model epistemic empathy, social caring, and comfort-building.

## ACKNOWLEDGMENTS

This project is supported by the National Science Foundation (NSF) grant, DRL 201088. We especially thank the participating teachers for their invaluable contributions. We also appreciate the anonymous reviewers who provided insights and expertise that assisted in the framing of the paper.

## ORCID

*Maggie S. Mahmood*  https://orcid.org/0000-0001-8451-6896
*Hamideh Talafian*  https://orcid.org/0000-0002-5094-6801
*Devyn Shafer*  https://orcid.org/0000-0001-8123-7525
*Eric Kuo*  https://orcid.org/0000-0001-5292-6188
*Morten Lundsgaard*  https://orcid.org/0000-0003-0668-4522
*Tim Stelzer*  https://orcid.org/0000-0001-6863-2045

## SUPPORTING INFORMATION

Additional supporting information can be found online in the Supporting Information section at the end of this article.